\documentclass[12pt]{article}
\newcommand{\bell}{\mathbf{\ell}}

\newcommand{\D}{\mathfrak{D}}

\newcommand{\bW}{{\mathbf{W}}}
\newcommand{\half}{{\textstyle \frac{1}{2}}}
\newcommand{\E}{{\mathcal{E}}}
\newcommand{\Eint}{{\mathcal{E}_{\rm int}}}

\newcommand{\bx}{\mathbf{x}}

\newcommand{\bb}{\mathbf{B}}
\newcommand{\bq}{\mathbf{q}}
\newcommand{\bv}{\mathbf{v}}
\newcommand{\bzero}{\mathbf{0}}
\newcommand{\Order}{\mathcal{O}}

\newcommand{\bel}[1]{\begin{equation}\label{#1}}
\newcommand{\bal}[1]{\begin{eqnarray}\label{#1}}
\newcommand{\ee}{\end{equation}}
\newcommand{\ea}{\end{eqnarray}}
\newcommand{\equ}[1]{~Eq.(\ref{#1})}

\newcommand{\Tr}{{\rm Tr}}
\renewcommand{\P}{\mathcal{P}}
\renewcommand{\H}{\mathfrak{H}}
\newcommand{\V}{\mathcal{V}}
\newcommand{\drop}[1]{}

\usepackage{graphicx}
\usepackage{latexsym}
\usepackage{amssymb}
\begin{document}

\vspace{0.5cm}
\begin{center}
{\bf Numerical Analysis of some Generalized Casimir Pistons}
\end{center}

\begin{center}
M. Schaden

Department of Physics, Rutgers University, Newark, NJ 07102
\end{center}
\begin{abstract}
The Casimir force due to a scalar field on a piston in a cylinder of radius $r$ with a spherical cap of radius $R>r$ is computed numerically
in the world-line approach. A geometrical subtraction scheme gives the finite interaction energy that determines the Casimir
force. The spectral function of convex domains is obtained from a probability measure on convex surfaces that is induced by the Wiener measure on
Brownian bridges the convex surfaces are the hulls of.  The vacuum force on the piston by a scalar field satisfying Dirichlet boundary
conditions is attractive in these geometries, but the strength and short-distance behavior of
the force depends crucially on the shape of the piston casing. For a cylindrical casing with a hemispherical head, the force
for $a/R\sim 0$ does not depend on the dimension of the casing and numerically approaches $\sim - 0.00326(4)\hbar c/a^2$.
Semiclassically this asymptotic force is due to short, closed and non-periodic trajectories that
reflect once off the piston near its periphery. The semiclassical estimate $-\hbar c/(96\pi a^2)(1+2\sqrt{R^2-r^2}/a)$ for the force
when $a/r\ll r/R\leq 1$ reproduces the numerical results within statistical errors.

\end{abstract}

PACS:

\vfill\eject
\section{Introduction}
Until fairly recently, only the Casimir force between parallel plates\cite{Casimir48,Lifshitz55} could be reliably
computed. Although the multiple scattering expansion of Balian, Bloch and Duplantier\cite{BB70,BD77} allows
for more complicated geometries, it was not used extensively to investigate the geometry dependence of Casimir forces beyond lowest order.
Renormalization issues further obscured the relationship between finite Casimir energies and physical
effects. Boyer\cite{Boyer68,Milton78} calculated the (finite and apparently repulsive) Casimir pressure on an
idealized metallic spherical shell in 1968 and DeRaad et al.\cite{DRM81} the pressure on a cylindrical one in 1981. It later
became clear that the finiteness of such Casimir self-energies depends rather sensitively on the assumed ultrathin ideal metallic boundaries\cite{Deutsch79,Graham01} and the physical interpretation of these results is debated.

In the past few years this situation has changed dramatically. A better understanding of the
relation between scattering theory and finite Casimir energies\cite{Klich06} allows one to in
principle compute Casimir forces between disjoint bodies with arbitrary optical characteristics. A
multipole expansion of the scattering matrix is particularly suited when the distance between the
bodies is large whereas a semiclassical approximation becomes appropriate and rather accurate at
short distances. The interaction vacuum energy of disjoint bodies is finite and the problem of
determining the Casimir force between them in this sense has been largely solved -- ignoring some issues
at finite temperature\cite{Mostepanenko08}.

Below I define finite subtracted vacuum (interaction) energies that determine
Casimir forces between bodies that are not separate. A similar geometrical approach was first used by
Power\cite{Power64} to obtain the well-known Casimir force between two parallel plates without
regularizing potentially infinite zero-point energies. Svaiter and Svaiter\cite{Svaiter92} emphasized that these Casimir pistons
show that certain (finite) physical vacuum effects do not depend on the regularization and renormalization procedure.
Recently there has been renewed interest in such geometries\cite{Cavalcanti04}
because the electromagnetic vacuum self-energy of a cube (as that of a sphere\cite{Boyer68}) is positive\cite{Lukosz71}.
The force on a partition in a parallelepiped nevertheless is attractive\cite{Jaffe05} at any position. For scalar
fields satisfying Dirichlet boundary conditions this is a consequence of reflection positivity\cite{Klich06,Bachas06}.

Semi-classical considerations suggest that the Casimir force on a piston strongly depends on the
shape of the casing. The vacuum force due to a massless scalar
field satisfying Dirichlet boundary conditions is here obtained numerically for some generalized Casimir pistons.
The world-line approach to Casimir interaction energies of Gies et al.\cite{GLM03} is thereby generalized to
geometries with connected boundaries. In Casimir pistons all domains are bounded and the
mathematical treatment in fact is much simpler and clear-cut than for the scattering situation with a continuous
spectrum. A finite subtracted vacuum energy gives rise to the force on the piston. In the
examples studied here, all domains are convex and numerical computations are vastly simplified by
considering the convex hulls of Brownian bridges.

\section{World-line approach to the interaction energy of connected bounded domains}
Consider the heat kernel operator $\mathfrak{K}_\D(\beta)=e^{{\beta\triangle}/2}$ for the Laplacian
$\triangle$ with Dirichlet boundary conditions on a bounded domain $\D\subset\mathbb{R}^3$. The
spectrum of eigenvalues  $\{\lambda_n>0,n\in\mathbb{N}\}$ of the negative Laplace operator in
this case is discrete, real and positive. The corresponding spectral function (or partition
function or trace of the heat kernel),
\bel{spfunc}
\phi_\D(\beta)=\Tr\mathfrak{K}_\D(\beta)= \sum_{n\in\mathbb{N}} e^{-\beta\lambda_n/2},
\ee
is finite for $\beta>0$. It has the well-known\cite{Greiner71,Gilkey84,Kirsten02,Fulling07b} high-temperature (short-time)
expansion,
\bel{assfunc}
\phi_\D(\beta\sim 0)\sim \frac{1}{(2\pi\beta)^{3/2}}\sum_{n=0}^\infty (2\pi\beta)^{n/2} a_n(\D)
+\Order(e^{-l^2/\beta}).
\ee
For smoothly bounded domains, the Hadamard-Minakshisundaram-DeWitt-Seeley coefficients $a_n(\D)$ in
this series are integrals over powers of the local curvature and reflect average geometric
properties of the domain and its boundary\cite{kac66,Vassilevich02}. For a bounded
three-dimensional flat Euclidean domain $\D$, $a_0(\D)$ gives its volume $\V_\D$ and
$a_1(\D)=-\mathcal{S_\D}/4$ gives the surface area $\mathcal{S_\D}$ of its boundary\cite{kac66}.
$a_2(\D)$ is proportional to the integrated curvature [sharp edges of the boundary also contribute
\cite{SW71,BD77}] and the dimensionless coefficient $a_3(\D)$ depends on topological
characteristics of the domain [such as the connectivity of its boundary and the number and opening
angles of its corners\cite{kac66,BD77}]. Since $a_4(\D)\neq 0$ implies a logarithmic divergent vacuum
energy that prevents a unique definition of the Casimir energy, this coefficient is crucial. Its geometric interpretation\cite{Vassilevich02} is, however, not easily explained. Non-analytic and (for $\beta\sim 0$) exponentially suppressed contributions to the asymptotic expansion of
$\phi_\D(\beta)$ are associated with classical periodic- and diffractive- orbits\cite{Brackbook} of minimal length $l$.

The world-line approach to Casimir energies\cite{GLM03} is based on the fact\cite{kac66,
Stroock93} that the spectral function for a bounded flat Euclidean domain $\D$ can be expressed in
terms of its support of standard Brownian bridges. In three dimensions,
\bel{support}
\phi_\D(\beta)=\int_\D \frac{d\bx}{(2\pi \beta)^{3/2}} \P[\bell_\beta(\bx)\subset\D ]\ ,
\ee
where $\bell_\beta(\bx)=\{\bb_\tau(\bx,\beta), 0\leq \tau\leq \beta;
\bb_0(\bx,\beta)=\bb_\beta(\bx,\beta)=\bx\}$ is a standard Brownian bridge from $\bx$ to $\bx$ in
"proper time" $\beta$ and $\P[\bell_\beta(\bx)\subset\D ]$ denotes the probability that such a bridge
is entirely within the bounded domain $\D$. Note that $\P[\bell_\beta(\bx)\subset\D ]/(2\pi \beta)^{3/2}$ is the Green function
from $\bx$ to $\bx$ in time $\beta$ of the associated diffusion problem with Dirichlet boundary conditions on $\partial\D$.

\equ{support} implies that the spectral function for a domain of finite volume is finite. Divergences arise only
in the corresponding zero-point energy. For finite vacuum energies, the leading (five, in three dimensions,)
coefficients in the asymptotic power series of\equ{assfunc} would have to vanish. Although impossible
for a single domain, one can improve the asymptotic behavior by considering a (finite) linear
combination of spectral functions for domains $\{\D_k;k=0,1,\dots,M\}$
\bel{tildephi}
\tilde\phi(\beta)=\sum_{k} c_k \phi_{\D_k}(\beta)\ .
\ee
When the coefficients $c_k$ are such that
\bel{finiteness}
\sum_{k} c_k a_i(\D_k)=0\ \ {\rm for}\ \ i=0,\dots,4 ,
\ee
the "interaction" vacuum energy,
\bel{inter}
\Eint=-\pi\int_0^\infty \frac{d\beta}{(2\pi\beta)^{3/2}}\tilde\phi(\beta)=\sum_{k} c_k \E_{\rm
vac}(\D_k)\ ,
\ee
is finite because the integrand is $\Order(\beta^{-1/2})$ for $\beta\sim 0$ when\equ{finiteness} holds.
The $\E_{\rm vac}(\D_k)\sim\half\sum_n \sqrt{\lambda_n}$ in \equ{inter} are the
(divergent) formal zero-point energies of a massless scalar field satisfying Dirichlet boundary
conditions on the individual bounded domains $\D_k$ . The linear combination $\Eint$ of these
vacuum energies may be interpreted as the difference in zero-point energy for domains of the same
total volume, total surface area, \emph{local} curvature, topology,etc... The subtractions have to be
chosen so as to at most affect the physical quantity of interest in
a calculable way. They thus could depend on the effect one wishes to describe.

In the context of Casimir energies, a similar subtraction scheme was first used by
Power\cite{Power64} to calculate the original Casimir force between parallel metallic
plates\cite{Casimir48} without regulators. Svaiter~\&~Svaiter\cite{Svaiter92} emphasized that
the subtracted vacuum energy of parallelepiped pistons is physical and does not depend on calculational procedures.
In reference \cite{Schaden06} non-analytic contributions to the finite interaction
vacuum energy were computed in leading semiclassical approximation. For numerical calculations, the subtraction method
outlined above is preferable to conventional regularization procedures, because we shall see that it can be
implemented by a restriction on the set of paths. One thus avoids the numerically difficult computation of small
differences in large quantities.

It is interesting (and will be useful numerically) to note that for \emph{convex} domains $\D$ a Brownian bridge $\bell_\beta(\bx)$ is wholly
within $\D$ only if its convex hull $\H[\bell_\beta(\bx)]$ is entirely contained in $\D$,
that is
\bel{hullcond}
\P[\bell_\beta(\bx)\subset\D ]=\P[\H[\bell_\beta(\bx)]\subset\D]\  {\rm
for}\ \D\ {\rm convex}\ .
\ee

In the continuum limit, a standard Brownian bridge process satisfies the stochastic differential
equation\cite{Stroock93,Oksendahl87},
\bel{SDE}
d\bb_\tau(\bx,\beta)=\frac{\bx-\bb_\tau(\bx,\beta)}{\beta-\tau}d\tau+d\bW_\tau,\ \ {\rm with}\ \
\bb_0(\bx,\beta)=\bx
\ee
where $\bW_\tau$ is a (3-dimensional) standard Wiener process, i.e. each component of $\bW_\tau$
is normally distributed about the origin $\bzero$ with variance $\tau$. Translational invariance and the
scaling property of random walks (and their convex hulls) imply that,
\bel{invariance}
\P[\bell_\beta(\bx)\subset\D
]=\P[\bell_1(\bzero)]\Theta[\bx+\sqrt{\beta}\bell_1(\bzero)\subset\D]=
\P[\bell_1(\bzero)]\Theta[\bx+\sqrt{\beta}\H[\bell_1(\bzero)]\subset\D]\ .
\ee
Here multiplication and addition operations distribute over all members of a point-set, that is they scale and translate the whole loop or hull.
Using \equ{invariance} to make the dependence on $\bx$ and $\beta$ explicit in \equ{support} and
inserting the result in \equ{inter} one obtains,
\bal{final}
\Eint&=&-\frac{1}{8\pi^2}\int d\bx \langle\int_0^\infty \frac{d\beta}{\beta^3} \sum_{k} c_k
\Theta[\bx+\sqrt{\beta}\H[\bell_1(\bzero)]\subset\D_k]\rangle_{\bell_1(\bzero)}\nonumber\\
&=&-\frac{1}{4\pi^2}\langle\int d\bq\int_0^\infty d\lambda \sum_{k} c_k
\Theta[\bq+\H[\bell_1(\bzero)]\subset\lambda \D_k]\rangle_{\bell_1(\bzero)}\ ,
\ea
where the expectation $\langle\dots\rangle_{\bell_1(\bzero)}$ is with respect to unit loops (standard Brownian
bridges over unit time). \equ{final} is the basic formula used for computing Casimir interaction energies in the world-line
approach\cite{GLM03}. Note that the overall proportionality constant in \equ{final} is $4$ times
that in\cite{GLM03} because the unit loops here are defined by a standard Brownian bridge process
rather than one of twice the variance [this can be verified by letting
$\beta\rightarrow 2 T$ in\equ{final} and comparing with\cite{GLM03}]. Although the straightforward interpretation of an integrated interaction
energy density is lost by the change in integration variables $\bx=\bq/\lambda, \beta=\lambda^{-2}$, the
last expression for $\Eint$ in\equ{final} is slightly better adapted to numerical evaluation.

\begin{figure}[ht]
\includegraphics[width=3.5in]{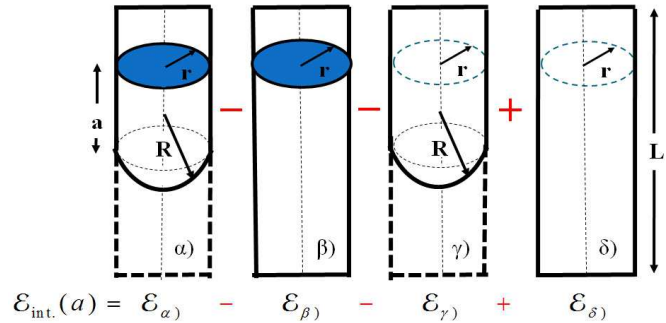}
\label{pistons}
\caption{\small The interaction Casimir energy $\Eint(a)$ for a cylindrical cavity of radius $r$ with a cap of radius $R\ge r$ and a moveable
planar piston at height $a$.  $\Eint(a)$ is the difference in vacuum energy of the cavity with and without piston compared to this
difference for a cylinder of the same radius $r$. This difference of differences in vacuum energies is finite for all values of
$a,r, L>2r$ and $R\geq r$. $\Eint$
does not require regularization and for $L\gg a$ only $\E_{\alpha)}$ depends on the position of the piston.
Solid lines denote surfaces on which Dirichlet boundary conditions are imposed. Thick dashed lines in Figs. $1\alpha)$ and $1\gamma)$ indicate
where the surface of the cylinder would be: Brownian bridges that pierce these surfaces give no contribution to $\Eint(a)$, leading
to condition $(\ref{condition})$ for large $L\gg 2 r$. Note that all five different bounded domains are \emph{convex} and bounded for finite $L$.}
\end{figure}

\section{Algorithm for the Casimir pistons of Fig.~1}
Fig.~1 depicts a combination of convex domains that is suitable for computing the interaction vacuum energy
$\Eint$ of a cylindrical Casimir piston of radius $r$ with a cap of radius $R\ge r$. A finite overall
length $L$ ensures that the volumes of all convex domains is finite. The restriction is of
little concern here because the limit $L\rightarrow\infty$ of $\Eint$ exists and in fact is easier to calculate.
Note that we could equally well have subtracted the vacuum energy at a fixed
position of the piston (as in \cite{Power64,Svaiter92,Boyer70} for the parallelepiped-piston),
but taking the limit $L\rightarrow\infty$ then would have been numerically challenging.

The main advantage of using the convex hull in numerical computations rather than the unit-loops,
is that the hull is a vastly reduced point-set that retains all the information on the
loop that is relevant for the partition function on convex domains [the sequential ordering and a great number of
interior points of the loop being discarded].
The number of vertices of the convex hull increases only logarithmically with the
number of points defining the loop. The convex hull of a loop given by a thousand points on average has
$\sim 55$ vertices and that of one with a million points has about $190$ vertices (see
Fig.~2). The time required by efficient algorithms to compute the hull (with $v$
vertices) of $n$ points is proportional to $n\log(v)$ \cite{Mount02}. In
evaluating the integrations in \equ{final} for sufficiently complex convex domains $\D$ like those
in Fig.~1, one only requires the convex hulls. For high precision, i.e.
for loops with a large number of points, $n$, it thus is advantageous to store the hulls rather
than the loops themselves. It would be preferable to generate convex surfaces \emph{directly}
with the appropriate measure, rather than to construct them as hulls of Brownian bridges.
But lacking a good algorithm to generate convex surfaces with the correct measure, they were here
constructed as hulls of Brownian bridges [systematic errors due to the finite number of
points of the original loops severely limit the accuracy one can achieve with this procedure].

\begin{figure}[ht]
\includegraphics[width=3.0in]{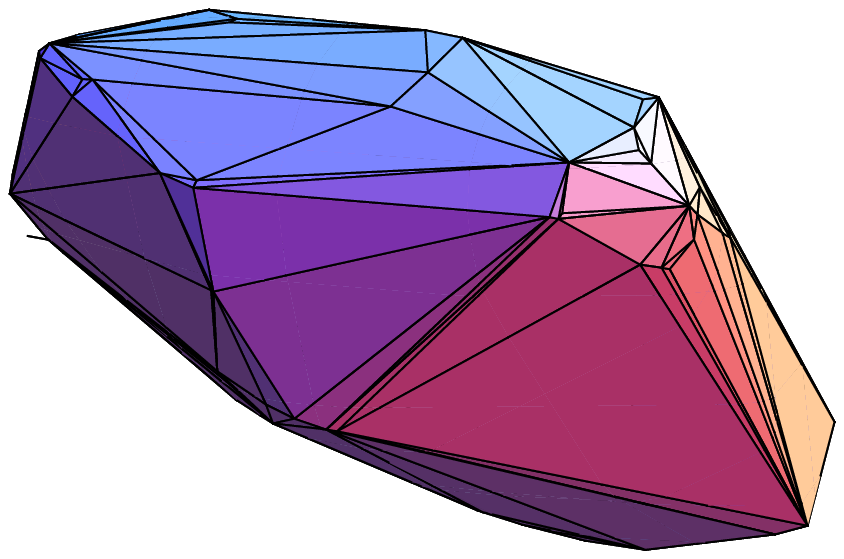}\includegraphics[width=3.5in]{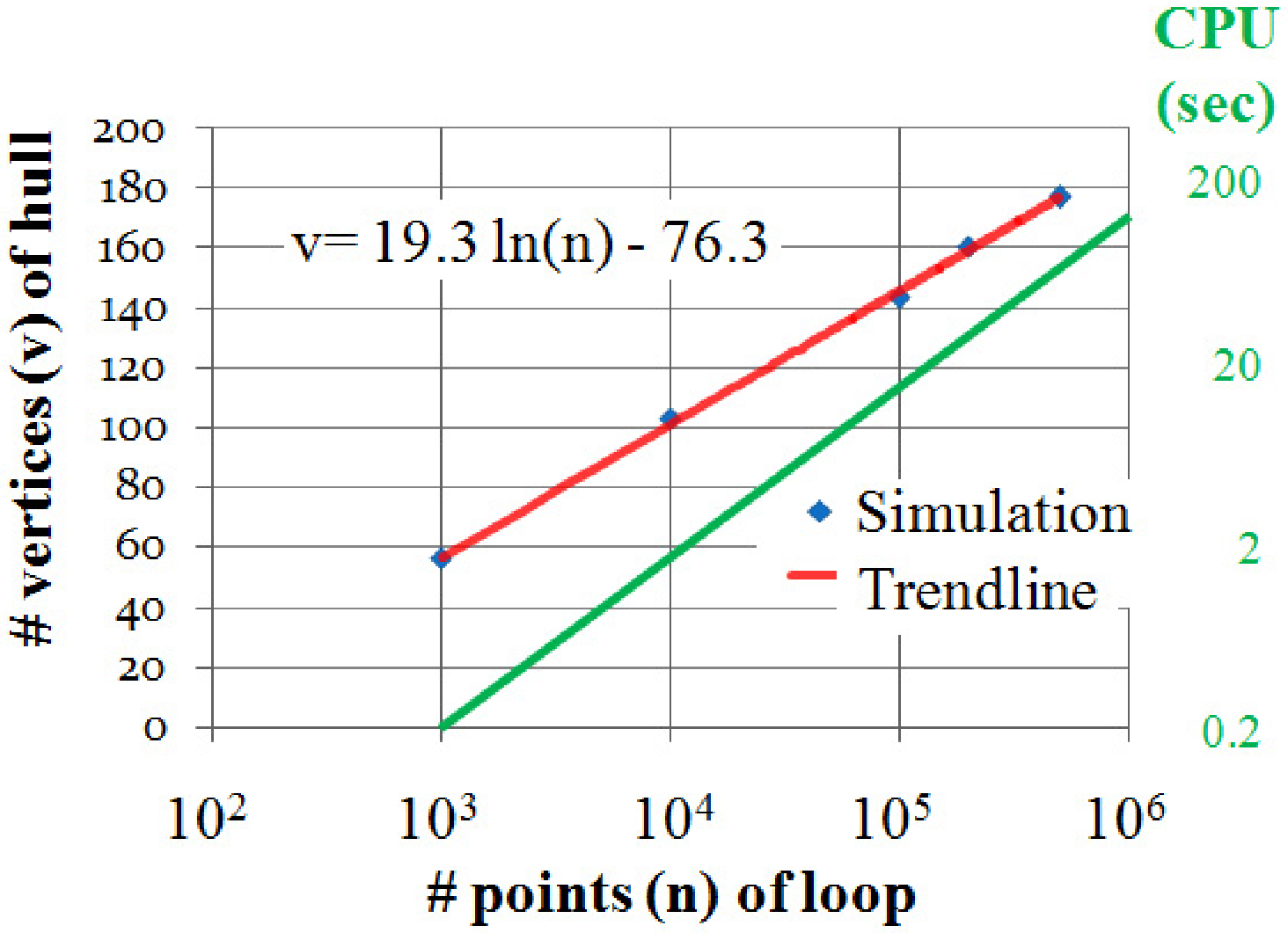}
\label{simulation}
\caption{\small On the left is a typical triangulated hull of a Brownian bridge defined by $10^5$ points. It has 286 faces and 145 vertices
and the shape of an irregularly cut rhinestone with large triangular facettes and intricate corners.  The average number of vertices (v) of
the convex hull of a Brownian bridge given by $n$ points can be read off the left axis of the graph using the upper (red) curve. Data points are
from numerical simulations. The inset is the formula for the (red) trendline that best describes the data. The right axis and lower (green)
line give the average CPU-time used to compute the hulls on a laptop with a 2GHz processor. Note that the left axis is linear while the
right axis is logarithmic.}
\end{figure}

To evaluate the spatial- and scale- integrals in \equ{final} for a given point-set
$\H[\bell_1(\bzero)]=\{\bv_i=(x_i,y_i,z_i);i=1,\dots,v\}$, note that (for $L\gg r$) the
only significant contribution to the interaction energy  $\Eint$ of the combination of domains shown in Fig.~1
comes from
\bel{condition}
\emph{loops (hulls) that pierce the piston and the piston head, but not the cylinder}.
\ee
The cylindrical symmetry of the domains in Fig.~1 allows one to rotate $\bq$ to the positive
$x$-axis and perform one angular integral trivially, resulting in
\bel{Episton}
\Eint({\rm cyl. piston})=-\frac{1}{2\pi}\langle\int_0^\infty \rho d\rho\int_{-\infty}^\infty dz
\int_0^\infty d\lambda \sum_{k} c_k \Theta[(\rho,0,z)+\H[\bell_1(\bzero)]\subset\lambda
\D_k]\rangle_{\bell_1(\bzero)}\ .
\ee
The remaining integrals have support in the region defined by,
\bel{region}
\begin{array}{ll}
a)&\exists\ i\in\{1,\dots,v\};\ z+z_i>\lambda a\ ,\ {\rm and}\\
b)&\forall\ i\in\{1,\dots,v\},\ (x_i+\rho)^2+y_i^2 <(\lambda r)^2\ ,\ {\rm and}\\
c)&\exists\ i\in\{1,\dots,v\};\  z+z_i<0\ \wedge\ (x_i+\rho)^2+y_i^2+(z+z_i-\lambda\sqrt{R^2-r^2})^2>(\lambda R)^2 \ .
\end{array}
\ee
Condition~\ref{region}a) implies that the convex hull pierces the piston at height $a$,
~\ref{region}b) that it does not pierce the cylinder of radius $r$ and~\ref{region}c)
that it pierces the cylinder head of radius $R\geq r$.

We first determine the maximum ($z_{\rm max}$) and minimum ($z_{\rm min}$) values of the set
$\{z_i;i=1,\dots,v\}$ for a convex hull. Given $\rho\in[0,(z_{\rm max}-z_{\rm min})r/a]$ and the
corresponding minimal value of $\lambda$ implied by \ref{region}b), the integration regions of
$\lambda$ and $z$ are then found by solving the non-linear optimization problem of \equ{region}
using a plane sweep algorithm\cite{Mount02} that determines intersections and endpoints of
the quadratic curves in \ref{region}c). Note that the quadratic functions of~\ref{region}c) decrease
monotonically for $z<-z_i$. Once the piecewise parabolic boundary of the integration region has
been obtained, the $z$- and $\lambda$-integrals of\equ{final} over this region are performed
analytically.

The remaining one-dimensional integral over $\rho$ has to be evaluated numerically. We compared two
methods: i) averaging the results of numerically integrating $\rho$ (using an adaptive algorithm)
for every hull separately and ii) averaging (scaled) values of the $\rho$-integrand of the hulls at gaussian
quadrature points and performing the numerical $\rho$-integration once only. For comparable
accuracy both methods require about the same computing time since most effort is spent in determining the integration region,
rather than in performing the integral: for comparable
accuracy, the adaptive algorithm on average requires less evaluations of the integrand than gaussian integration.

To verify the accuracy of the algorithm and estimate systematic errors, we compare (see Fig.~3) the
numerical results for $R\gg r\gg d$ with the known analytic Casimir energy due to a scalar
satisfying Dirichlet boundary conditions on two parallel circular plates of area $\mathcal{S}=\pi
r^2$ (half the electromagnetic Casimir energy\cite{Casimir48} of this configuration). For
$R\gtrsim r\gg a$ the numerical results are compared with the analytical estimates of the
asymptotic behavior discussed below.

\begin{figure}[ht]
\includegraphics[width=5in]{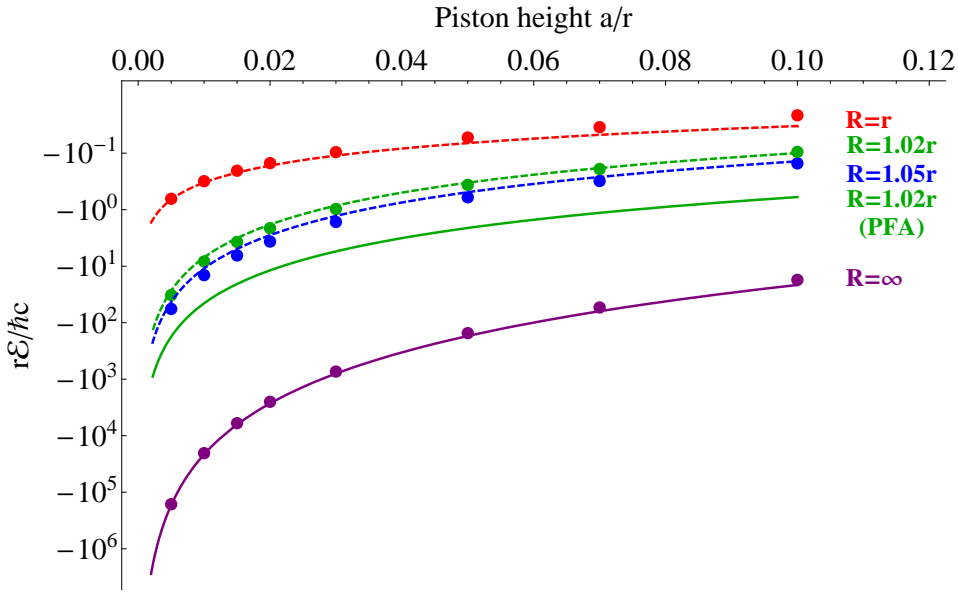}
\label{results}
\caption{\small The dimensionless Casimir interaction energy $r\Eint(a)/(\hbar c)$ for the cylindrical cavities of radius $r$ with caps of
radius $R\ge r$ of Fig.~1 as a function of the rescaled height $a/r$ of the moveable piston. Dots indicate numerical results
(dot size does not represent errors, which are too small to show on this logarithmic plot).
Solid lines are piston-based Proximity Force Approximation (PFA) estimates (note that the PFA overestimates the interaction energy by an order of
magnitude for $R=1.02r$). Dashed curves correspond to the force given by the asymptotic semiclassical estimate of \equ{Fline}.
Note the many orders of magnitude between a piston with a flat (bottom, purple) and a hemispherical (top, red) cylinder head.}
\end{figure}

\section{Estimate of the Casimir energy of a cylindrical piston with spherical cylinder head for $a/r\ll r/R\lesssim 1$}
\subsection{Direct heuristic approach}
The asymptotic contribution to the Casimir energy of the cylindrical piston in Fig.~1 for small
$a/r\ll 1$ is dominated by short loops of length $\Order(a)$ that satisfy\equ{region}. Since a unit
loop generally has extension in $z$-direction of $\Delta z=\Order(1)$, the main contribution when
$a/r\ll r/R$ comes from the behavior of the integrand of\equ{final} at large $\lambda\sim \Delta
z/a$  (small $\beta$) and for $r/R\sim 1$ is concentrated near $\rho\sim\lambda r\gg x_i,y_i$ for
all vertices. In this regime the boundary of the integration region in\equ{region} approaches,
\bel{asyregion}
\begin{array}{lrl}
a)&\Delta z=z_{\rm max}-z_{\rm min}\geq z_{\rm max}+z &>\lambda a \\
b)&\rho &<\lambda r\\
c)&\rho^2+(z+z_{\rm min}-\lambda\sqrt{R^2-r^2})^2&>(\lambda R)^2 \ ,
\end{array}
\ee
where $z_{\rm max}$ and $z_{\rm min}$ are the maximal/minimal $z$-values of the vertices of the
unit loop. \equ{asyregion} apparently is invariant under rotations of a hull about the z-axis, but
only because it was implicitly assumed that for any given $\lambda$ the hull is oriented
so that conditions~(\ref{region}) are satisfied for the largest possible value of $\rho$. As can be
seen from Fig.~4, this implies choosing a particular hull vertex as the origin of the hull and
orienting the hull optimally with respect to the casing by rotating it about its z-axis. Selecting just
one particular representative of a class of rotated hulls turns out to be a reasonable approximation for
obtaining the asymptotic behavior when $r\sim R $. [When $r=R$, the fact that the smallest possible
hull of a given shape must not pierce the cylinder wall while piercing the piston and hemispherical
cap evidently greatly constrains the orientation of the hull with the greatest weight -- slightly
rotating away from the optimal orientation requires smaller $\rho$ and $\lambda$ or may even make
it impossible to satisfy~(\ref{condition})]. Note that conditioning on hulls with a particular
orientation implies that the volume $V_{SO(2)}=2\pi$ of rotations about the z-axis should be
divided out of the rotation invariant probability measure for the hulls.
\begin{figure}[ht]
\includegraphics[width=2in]{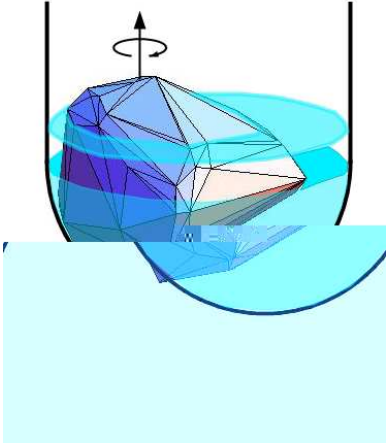}
\label{orientedhull}
\caption{\small An optimally oriented convex hull of a unit loop. The hull is rotated about a vertical axis
to maximize its contribution to $\Eint(a)$ in\equ{Episton} for small piston height $a$.  Almost all hulls have the shape of an irregularly
cut rhinestone (the one depicted here has 55 vertices and is that of a Brownian bridge defined by $10^3$ points). There
generally is just one optimal relative orientation of the hull with the greatest weight. Rotation symmetry about the cylinder axis has already been taken into account in\equ{Episton}. Note that for small $a$ and a hemispherical piston, a less optimal orientation leads to a great
reduction of the available phase space, since the hull must satisfy condition~(\ref{condition}) and
pierce the piston and cap but not the cylinder.}
\end{figure}
Requirements \ref{asyregion}a) and \ref{asyregion}c) imply that $\lambda a-z_{\rm max}\leq z\leq
\lambda \sqrt{R^2-r^2}-\sqrt{(\lambda R)^2-\rho^2}-z_{\rm min}$. The upper bound of the
$\rho$-integration is $\lambda r$ and its lower bound is $\sqrt{(\lambda R)^2 -(\lambda
\sqrt{R^2-r^2}+\Delta z-\lambda a)^2}$ when $\lambda R \geq\lambda \sqrt{R^2-r^2}+\Delta z-\lambda
a$, and zero otherwise. Neglecting terms that manifestly are finite in the limit $a/r\rightarrow
0$, the asymptotic behavior of $\Eint$ for $a/r\ll r/R$ (upon substituting $\rho^2=\lambda\sigma,\
\lambda=\Delta z/\gamma$) is given by,
\bal{finalasymp}
\Eint({\textstyle{\frac a r}\ll {\frac r R}})&\sim &-\frac{1}{4\pi}\frac{\langle(\Delta
z)^4\rangle_{\bell(0,1)}}{V_{SO(2)}}
\int^{\hspace{-0em}^{(a+R-\sqrt{R^2-r^2})}}_{_a}\hspace{-4em}\gamma^{-5} d\gamma\hspace{1em}
\int^{^{ r^2}}_{_{R^2 -(\sqrt{R^2-r^2}+\gamma - a)^2}} \hspace{-6em}(
\sqrt{R^2-r^2}-\sqrt{R^2-\sigma}-a +\gamma) d\sigma\nonumber\\
&=&-\frac{1}{96\pi^2 a}\left(\frac{\sqrt{R^2-r^2}}{a}+1+\Order(a/r)\right)\langle (\Delta
z)^4\rangle_{\bell_1(\bzero)}\ .
\ea

The expectation $\langle (\Delta z)^4\rangle_{\bell_1(\bzero)}$ is related to the Casimir
interaction energy of two parallel discs of radius $r$, i.e. the limit $r/R\ll a/r\ll 1$ of the
piston in Fig.~1. The integration region (\ref{region}) in this case is readily found and one
obtains,
\bal{flatE}
\Eint({\textstyle 0\sim{\frac r R}\ll{\frac a r}\ll 1})&\sim
&-\frac{1}{4\pi}\langle\int_0^{^{\Delta z/a}} d\lambda \int_0^{^{(\lambda r)^2}}(\Delta z-\lambda a) d\rho^2\rangle_{\bell_1(\bzero)}\nonumber\\
&=&-\frac{\pi r^2}{48\pi^2 a^3}\left(1+\Order(a/r)\right)\langle (\Delta z)^4\rangle_{\bell_1(\bzero)}\
.
\ea
Note that we have not divided by $V_{SO(2)}=2\pi$ in\equ{flatE} because the contribution from loops near the
cylinder surface is negligible in this limit and no particular orientation generally is much preferred. Comparing to
the known Casimir interaction energy due to a scalar field for two parallel discs of radius $r$ one has that,
\bel{expectation}
\langle (\Delta z)^4\rangle_{\bell_1(\bzero)}=\frac{\pi^4}{30}\ ,
\ee
a result for unit Brownian bridges that can also be found directly. For $a/r\ll
r/R$ the interaction energy of a Casimir piston in Fig.~1 with a cylinder head of radius $R\sim r$
thus is estimated to asymptotically approach,
\bel{cylasymptotic}
\Eint({\textstyle{\frac a r}\ll {\frac r R}\lesssim 1})=-\frac{\hbar c\pi^2}{2880
a}\left(\frac{\sqrt{R^2-r^2}}{a}+1+\Order(a/r)\right)\ .
\ee

Including only statistical errors, the asymptotic numerical data for $r\leq R\leq 1.02 r$ is best
reproduced by,
\bel{numasymp}
\Eint({\textstyle{\frac a r}\ll {\frac r R}\lesssim 1})\sim-\hbar
c\left(\frac{0.00395(5)\sqrt{R^2-r^2}}{a^2}+\frac{0.00326(4)}{a}+\Order(a/r)\right)\ .
\ee
The coefficients of the $\sqrt{R^2-r^2}/a^2$ and $1/a$ terms indeed are comparable and the leading
$0.00326(4)/a^2$-behavior of the attractive force on the piston with a hemispherical head  differs
by less than $5\%$ from the estimate obtained from\equ{cylasymptotic}. A semiclassical determination of the
asymptotic behavior gives a coefficient that is even closer to the one obtained numerically.  The semiclassical analysis below
furthermore shows that the asymptotic behavior of $\Eint$ is the same in magnitude but of
opposite sign for a scalar field satisfying Neumann boundary conditions.
\begin{figure}[ht]
\includegraphics[width=5in]{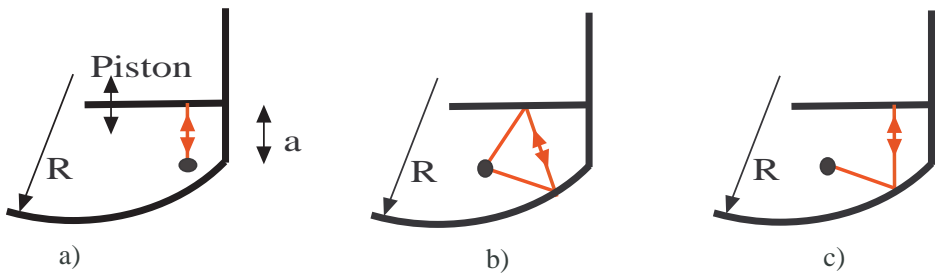}
\label{asymptotic5}
\caption{\small Some closed classical paths that depend on the piston height $a$ and become arbitrary short as $a\rightarrow 0$. a) closed paths with
just one reflection, b) paths with two reflections, c) paths with three reflections; similar closed paths with more reflections are not shown.
As explained in the main text only paths of type a) with just one reflection contribute to the asymptotic interaction energy. Contributions
due to  paths b) and c) vanish for Neumann and Dirichlet boundary conditions (in 3 dimensions).}
\end{figure}
\subsection{Semiclassical approach}
Semiclassically a Casimir energy diverges for some values of the geometric parameters due to closed
classical paths that become arbitrary short in this limit. The contribution of periodic orbits to
$\Eint(a=0)$ of a hemispherical piston for a scalar field satisfying Dirichlet boundary conditions
was obtained in\cite{Mateescu07},
\bel{dpistonperiodic}
\E_{\rm int}^{\rm per.}(a=0)=\frac{\hbar c\pi}{128 R}\left(1+
\frac{\pi^2}{45}+\sum_{m=1}^{\infty}\sum_{n=2m+1}^\infty\frac{30\sqrt{2}\cos(m\pi/n)}{
n^4\pi\sin^2(m\pi/n)}\right)\sim 0.044239\dots \hbar c/R\ .
\ee
The length of any periodic orbit of the hemisphere exceeds $2 r$ and the interaction Casimir energy
due to periodic orbits therefore is finite even at $a=0$.  However, for Dirichlet (or Neumann)
boundary conditions there are additional contributions to $\Eint$ from non-periodic closed
classical paths that depend on the position of the piston and become arbitrary short as
$a\rightarrow 0$. The closed paths shown in Fig.~5 potentially give a divergent contribution to the
Casimir energy in the $a\rightarrow 0$ limit. The trajectories of Fig.5b) and Fig.5c) have an odd number of conjugate points and
do not contribute\cite{Schaden06}.
Independent of this phase cancelation, the length of the trajectories in Fig.5b) and Fig.5c) increases rapidly with
$R$ and they therefore do not contribute to the Casimir energy for
$R\rightarrow\infty$. The paths of Fig.5b) and 5c) thus cannot possibly explain the heuristic
and numerical result that the Casimir energy of a hemispherical piston does not vanish for
$r=R\rightarrow\infty$. However, contrary to the electromagnetic case\cite{BD77}, the classical
trajectory of Fig.5a) that reflects just once off the piston for Dirichlet (D) or Neumann (N) boundary conditions
gives rise to the asymptotic force,
\bal{Fline}
\mathcal{F}_{D/N}^{\rm Fig.~3a)}=-\frac{\partial}{\partial a} \E^{\rm Fig.3a)}&=&\mp\frac{\hbar
c}{16\pi}\int_0^r\frac{\rho d\rho}{(a+\sqrt{R^2-\rho^2}-\sqrt{R^2-r^2})^4}\nonumber\\
&=&\mp\frac{\hbar c}{96\pi a^2}\left(\frac{2\sqrt{R^2-r^2}}{a}+1+\Order((a/r)^2)\right)\ .
\ea
Only terms that diverge for $a\rightarrow 0$ have been retained in this estimate, because other
non-singular contributions to the force of $\Order(1/r^2)$ have also not been included. Apart from a
3\% smaller overall strength ($\frac{\pi^2/2880}{1/(96\pi)}=1.033\dots$), the force in\equ{Fline}
is consistent with the asymptotic behavior in\equ{cylasymptotic} that was obtained heuristically. The
semiclassical calculation in particular also shows that the asymptotic $1/a^3$ behavior of the
force vanishes like $\sqrt{R^2-r^2}$ as $R\rightarrow r$. For a hemispherical Casimir piston, the
force for $a\ll r$ is proportional to $1/a^2$ and (somewhat surprisingly) does not depend on the
radius $r=R$ of casing and cylinder head. Note that the semiclassical coefficient $1/(96\pi)\sim
0.00331\dots$  of this asymptotic force is only 2\% larger than the coefficient $0.00326(4)$
found numerically. The 2\% discrepancy probably is as good an estimate of the systematic errors
of the numerical calculation as any one can give at this level of accuracy (the statistical error of
the calculations is about 1.2\%). Fig.~6 shows the \emph{difference} between the numerical results
and the asymptotic semiclassical energy for the hemispherical Casimir piston in a linear plot. Note
that the residual force is small and repulsive.
\begin{figure}[ht]
\includegraphics[width=5in]{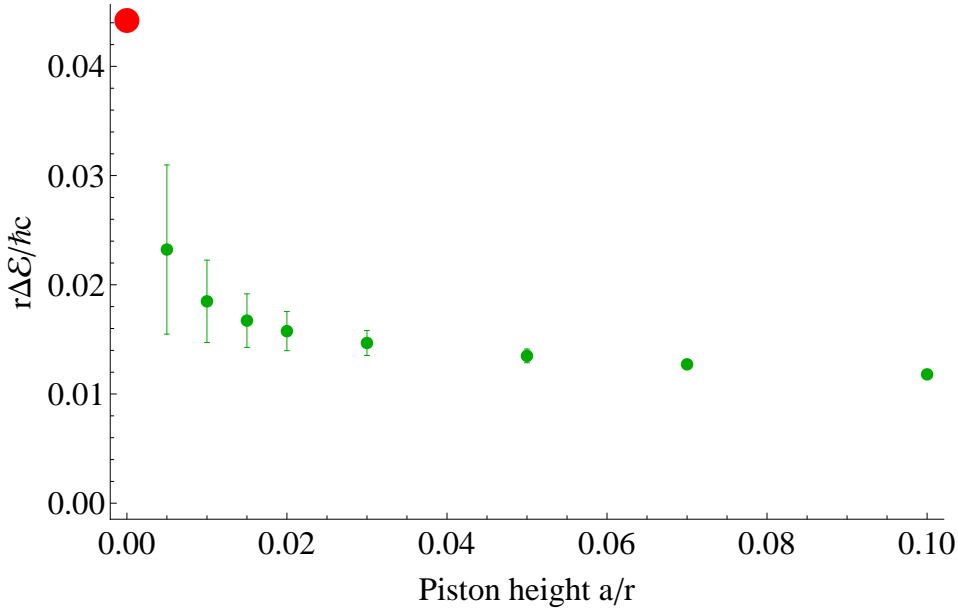}
\label{Difference6}
\caption{\small The difference $\Delta\E(a)=\Eint(a)+\hbar c/(96\pi a)$ of the numerical- and asymptotic semiclassical- estimate for the
scalar Casimir interaction energy of a hemispherical piston. The energy
difference is plotted in dimensionless units against the piston height $a$ in units of the cylinder radius $r$. Only statistical errors of
the numerical calculation are shown. The numerical results are those of Fig.~3 for $R=r$ and were obtained from the convex hulls
of $10^5$ unit loops, each given by $10^5$ points. The (red) dot at $a/r=0$ is the semiclassical contribution to this difference
due to periodic orbits given by\equ{dpistonperiodic}.}
\end{figure}

\section{Discussion and Conclusions}
The Casimir force on a piston by a massless scalar field satisfying Dirichlet boundary
conditions depends qualitatively on the shape of the casing in the systems of Fig.~1. Whereas the
attractive force for small piston height $a/r\ll 1$ is proportional to $r^2/a^4$ for a flat
cylinder head, it is proportional to $\sqrt{R^2-r^2}/a^3$ for heads with curvature radius $r< R \ll
r^2/a$ and for a hemispherical piston with ($R=r$) is proportional to $1/a^2$. Note that the asymptotic
behavior of \equ{Fline} implies that one percent deviation in radius from a hemispherical cylinder head
doubles the Casimir force on the piston at an elevation $a/r\sim 0.1$. The fact that the attractive
Casimir force is greatly reduced for a hemispherical cap is well born out by the numerical
calculations. Fig.~3 indicates that the Casimir force on the piston depends qualitatively on $R/r$ and at
small piston height $a/r$ may differ by several orders in magnitude. [Because a
larger cap radius implies a greater "average" separation between the piston and the cap, the
overall change in magnitude is somewhat deceptive, but qualitative changes in the Casimir force in
this logarithmic plot appear as a change in slope  and the reduction of the Casimir force in a
hemispherical piston clearly is not simply due to the increase in average piston height.]

The interaction Casimir energy defined in\equ{final} is \emph{a priori} finite (and does not
require regularization) when the conditions of\equ{finiteness} are satisfied. Only a portion of the
vacuum energy is computed that includes all its dependence on the piston height $a$ and determines
the force on the piston. To relate this interaction vacuum energy of a massless scalar
to the spectral function of the Laplace operator, leading terms in the high-temperature expansion
have to be canceled. Here this was achieved by subtracting spectral functions of domains with the
global characteristics represented by the first few (five in three dimensions) heat kernel
coefficients. The subtraction procedure in particular implies that finite Casimir energies are differences
in the vacuum energy of domains with the same topology, same volume, same average curvature etc...

The linear combination of vacuum energies for $\Eint$ in the present examples is shown in Fig.~1 -- for
$L\gg r$ only  $\E_{\alpha)}$ depends on the height $a$ of the piston. The Casimir energy due to a
massless Dirichlet scalar in this case is given by the positive probability measure for
Brownian paths that satisfy\equ{condition}. This implies an attractive Casimir force on the piston
for any separation $a$ and cap radius $R>r$. The same argument implies that the Casimir force by a Dirichlet scalar
is attractive for a large class of piston geometries for which a condition on the Brownian
bridges like \equ{condition} holds -- this includes pistons and cylinders of arbitrary cross-section and form with caps
that are entirely within the cylinder. Cylinder heads that extend beyond the cylinder have also been examined\cite{Schaden08a}.

For convex Euclidean domains, the spectral function is related to a probability measure on convex surfaces.
This measure is indirectly induced by that on Brownian bridges the convex surfaces are the hulls of.
Convex surfaces in this investigation were constructed from the Brownian loops and were not generated
directly. It would be very advantageous for the accuracy and speed of numerical simulations to
directly construct convex surfaces with the appropriate measure.

A semiclassical analysis of the geometries presented here reproduces the numerical results for the force on
the piston to better than 2\% when $a/r<0.1$, it in particular gives the correct asymptotic behavior
of the force for $a/r\rightarrow 0$ whereas the PFA is inapplicable and overestimates the force by an order of magnitude.
For a massless scalar satisfying Dirichlet boundary
conditions, this asymptotic behavior is semiclassically given by the contribution of the single closed
classical path of Fig.~3a) that reflects just once off the piston. The contribution to the Casimir
energy of \emph{periodic} orbits calculated in\cite{Mateescu07} is finite at $a=0$ and negligible for
$a/r\ll r/R$. The non-periodic path of Fig.~3a) determines the asymptotic behavior of the Casimir
energy because its length is $\Order(a)$ near the edge of the piston, whereas all periodic paths have a length exceeding
$2(R-\sqrt{R^2-r^2})$. For a hemispherical cap the attractive Casimir force on the
piston $-\hbar c/(96\pi a^2)$ does not depend on the radius $r=R$ of the cylinder and cap and
mimics the electrostatic force on a metallic piston a charge $q^2/(\hbar c)=1/(48\pi)\sim 1/150.8
\lesssim\alpha_{EM}$ at the center of the spherical cap would exert. Although the same as
the force due to a charge, the Casimir force here is inherently non-local with an energy density that is
greatest near the intersection of cylinder and cap. The (non-local) nature of this force becomes
drastically apparent if one lets the cylinder- and cap- radius tend to infinity while keeping the
height $a$ of the piston constant: the Casimir force on the piston (locally a flat plate) remains
unchanged in this limit and the direction of the force would give information about the position
of the piston relative to the cap at infinity. This paradoxical situation arises only for the
Casimir energy of a massless scalar field satisfying Dirichlet (Neumann) boundary conditions. Semiclassically
the asymptotic force due to a scalar field satisfying Neumann boundary conditions in\equ{Fline} is of the same magnitude but of opposite sign.
[In the method of images the "mirror source" for Neumann boundary conditions at the piston is of the \emph{same} sign as the source, whereas it is of opposite sign for Dirichlet boundary conditions].

The semiclassical calculation in\cite{Mateescu07} suggests that the electromagnetic Casimir force on a piston with hemispherical cap vanishes for $r\rightarrow \infty$ and is weakly repulsive. Because closed paths with an odd number of reflections vanish in the electromagnetic case\cite{BD77}, one does not expect the electromagnetic interaction energy of a hemispherical piston to diverge\cite{comment1} for $a\rightarrow 0$.
Indeed, if the electromagnetic Casimir force for small $a/r$ is $\Order(1/r^2)$, one cannot deduce the relative orientation and
position of cap and piston for $r\rightarrow\infty$. [One does not encounter this problem with locality for two flat
plates: for $a\ll r$ the distance between the plates and their relative orientation can always be found
locally and does not require knowledge of the situation at distances $\Order(r)$ from the center.] In contrast to Dirichlet boundary conditions, a repulsive force on a piston in a metallic cylinder does not violate reflection positivity, because the metallic cylinder induces correlations in the electromagnetic fluctuations on either side of the piston.
That vacuum forces in some geometries are small or at least comparable to electrostatic forces down to very
small separations could have implications for precision measurements of forces and the design of Micro-Electro-Mechanical devices.

\noindent{\bf Acknowledgements:} I would like to thank the organizers of the International Workshop "60 Years of Casimir Effect" in Brasilia for a very engaging conference and I.~Klich for some illuminating discussions on the implications of reflection positivity. This invesigation was supported by
the National Science Foundation with Grant No. PHY-0555580.


\begin{thebibliography}{99}


\bibitem {Casimir48} Casimir H.B.G.,
% On the attraction between two perfectly conducting plates
 {\it Proc. Kon. Nederland. Akad. Wetensch.} {\bf B51}, 793 (1948).

\bibitem{Lifshitz55}
E.M. Lifshitz, Zh. E´ksp. Teor. Fiz. {\bf 29}, 94 (1955) and Sov. Phys. JETP
2, 73 (1956); I.E. Dzyaloshinskii, E.M. Lifshitz, and l. Pitaevskii, Adv.
Phys. {\bf 10}, 165 (1961).
%Dielectrics %%

\bibitem{BB70} R.~B.~Balian and C.~Bloch, Ann. Phys. (NY) {\bf
60}, 401 (1970);{\it ibid} {\bf 63}, 592 (1971);{\it ibid} {\bf 64}, 271 (1971) Errata in {\it ibid} {\bf
84},559 (1974); {\it ibid} {\bf 69}, 76 (1972);{\it ibid} {\bf 85}, 514 (1974).

\bibitem{BD77} R.~Balian and B.~Duplantier,
 %``Electromagnetic Waves Near Perfect Conductors. 1. Multiple Scattering
%Expansions. Distribution Of Modes,''
Annals Phys.\  {\bf 104}, 300 (1977); ibid  {\bf 112}, 165 (1978);{\it ibid}, Proceedings of the
15$^{\rm th}$ SIGRAV Conference on General Relativity and Gravitational Physics held at Villa
Mondragone, Monte Porzio Catone,Italy, September 9-12 (2002); quant-ph/0408124.

\bibitem{Boyer68} T.~H.~Boyer,
%``Quantum Electromagnetic Zero Point Energy Of A Conducting Spherical Shell
%And The Casimir Model For A Charged Particle,''
Phys.\ Rev.\  {\bf 174}, 1764 (1968).
%%CITATION = PHRVA,174,1764;%%

\bibitem{Milton78} K.~A.~Milton, L.~L.~DeRaad and J.~S.~Schwinger,
%``Casimir Selfstress On A Perfectly Conducting Spherical Shell,''
Annals Phys.\  {\bf 115}, 388 (1978).
%%CITATION = APNYA,115,388;%%

\bibitem{DRM81} l.~L.~DeRaad, Jr. and K.~A.~Milton, Ann. Phys. (N.Y.) \textbf{136}, 229 (1981).
%first cylinder calculation

\bibitem{Deutsch79} D.~Deutsch and P.~Candelas, Phys. Rev. {\bf D 20}, 3063 (1979);P.~Candelas,
    Ann.Phys. (NY) {\bf 143}, 241 (1982);{\it ibid},{\bf 167}, 257 (1986).

\bibitem{Graham01} N. Graham, R. L. Jaffe, V. Khemani, M. Quandt, M. Scandurra and H.
Weigel, Nucl.Phys. B{\bf 645} (2002) 49; N. Graham, R. L. Jaffe, M. Quandt, O. Schr\"{o}der, and H.
Weigel, Nucl. Phys. B{\bf 677}, 379 (2004).

\bibitem{Klich06} O.~Kenneth and I.~Klich,
% Opposites Attract - A Theorem about the Casimir Force,
Phys. Rev. Lett. {\bf 97}, 160401 (2006).

\bibitem{Mostepanenko08}  G. L. Klimchitskaya and V. M. Mostepanenko, J. Phys. A {\bf 41}, 312002(F) (2008);
S. K. Lamoreaux, [arXiv:0801.1283]; R. S. Decca, E. Fischbach, B. Geyer, G. L. Klimchitskaya,
D. E. Krause,D. Lopez, U. Mohideen, and V. M. Mostepanenko, [arXiv:0803.4247]

\bibitem{Power64} E.~A.~Power, \emph{Introductory Quantum Electrodynamics} (Elsevier, New York,
    1964), Appendix I.
\bibitem{Svaiter92} N.F.~Svaiter and B.F.Svaiter, J.~Phys. A {\bf 25},979 1992.

\bibitem{Cavalcanti04} R.M. Cavalcanti, Phys. Rev. D {\bf 69},065015 (2004).
%Casimir Piston

\bibitem{Lukosz71} W.~Lukosz, Physica {\bf 56}, 109 (1971).

\bibitem{Jaffe05} M.P.~Hertzberg, R.L.~Jaffe, M.~Kardar, A.~Scardicchio, Phys. Rev. Lett. {\bf 95},
    250402 (2005);{\it ibid},Phys.Rev.D {\bf 76}, 045016 (2007).

\bibitem{Bachas06} C.P.~Bachas, J.Phys.A{\bf 40},9089,2007.

\bibitem{GLM03} H.~Gies and K.~Langfeld, Int.J.Mod.Phys. A \textbf{17}, 966 (2002);
    [hep-th/0112198]
    H.~Gies, K.~Langfeld and L.~Moyaerts, JHEP {\bf 0306}, 018
    (2003);
    %[hep-th/0303264]
    \emph{ibid}, Proceedings of QFEXT'03 (Norman, Sept. 2003) 203 [hep-th/0311168];
     H.~Gies and K.~Klingmuller, Proceedings
    of QFEXT'05 (Barcelona, Sept. 2005)  [hep-th/0511092];
    \emph{ibid}, Phys.~Rev.~{\bf D74}, 045002 (2006);
    % [quant-ph/0601094]
    See C. Schubert, Phys. Rept. 355 (2001) 73 for a
    review of the worldline formalism.

\bibitem {Greiner71} Greiner P.,
% An asymptotic expansion for the heat equation.,
 {\it Arch Rat Mech Anal}, (1971), {\bf 41}, 163–218

\bibitem {Gilkey84} P.B. Gilkey,
 Invariance Theory, the Heat Equation, and the Atiyah-Singer Index Theorem,
 (Publish or Perish, Wilmington,1984) and (CRC Press, Boca Raton 1995).

\bibitem {Kirsten02} K. Kirsten, Spectral functions in Mathematics and Physics, (Chapman \& Hall/CRC Press, Boca Raton 2002)

\bibitem{Fulling07b} S.A. Fulling,
% Vacuum Energy as Spectral Geometry,
SIGMA {\bf 3}, 094 (2007).

\bibitem{kac66} Kac M.,
% Can one hear the shape of a drum?,
 {\it Amer. Math. Monthly} {\bf 73} Part II, 1 (1966).

\bibitem{Vassilevich02} D.V. Vassilevich, Physics Rep. 388 (2003) 279 and references therein.

\bibitem{SW71} Stewartson K, Waechter R T.
% On hearing the shape of a drum: further results.
 {\it Proc. Camb. Philos. Soc.}{\bf 69} (1971)  353.

\bibitem{Brackbook} M.~Brack and R.~Bhaduri, \emph{Semiclassical Physics} (Addison-Wesley, Reading
    1997).


\bibitem{Stroock93} D.W. Stroock, Probability Theory, an Analytic View (Cambridge Univ. Press.,
    Cambridge 1993).

\bibitem {Schaden06} M. Schaden,
% Comments on the Sign and Other Aspects of Semiclassical Casimir Energies,
Phys. Rev. A {\bf 73} (2006) 042102.
% [arXiv:hep-th/0509124v2].

\bibitem{Oksendahl87} B. {\O}ksendahl,\emph{Stochastic Differential Equations} 5$^{\rm th}$
    Ed. (Springer, 2000), p.75; see also L.C.G. Rogers and D. Williams, \emph{Diffusions,
    Markov Processes, and Martingales. Vol. 2} (J. Wiley \& Sons, 1987).

\bibitem{Boyer70} H. Boyer, Ann. Phys. (N.Y.) 56, 474 (1970).

\bibitem{Mount02} D.M. Mount, lecture notes for course CMSC 754 \emph{Computational Geometry}
    (Univ. of Maryland, 2002).

\bibitem {Mateescu07} M. Schaden and L. Mateescu,
\emph{Weak and Repulsive Casimir Force in Piston Geometries},
 [arXiv:0705.3435v1].

\bibitem{Schaden08a} M. Schaden, \emph{On the Direction of Casimir Forces}, [arXiv:0808.3966] .

\bibitem{comment1} A for $a\rightarrow 0$ divergent force can only arise due to
high-frequency modes that should be well described semiclassically. At high frequencies
transverse electric (TE) and transverse magnetic (TM) modes furthermore satisfy
Dirichlet and Neumann boundary conditions respectively.







\drop{
\bibitem{Experiments}
S.~K.~Lamoreaux, %"Demonstration of the Casimir Force in the 0.6 to 6 µm Range"
Phys. Rev. Lett. {\bf 78}, 5 (1997); %
U.~Mohideen and A.~Roy, Phys. Rev. Lett. {\bf 81}, 4549 (1998); T.~Ederth, Phys. Rev. {\bf A 62},
062104 (2000); G.~Bressi, G.~Carugno, R.~Onofrio and G.~Ruoso, Phys. Rev. Lett. {\bf 88}, 041804
(2002); R.~S.~Decca, D.~Lopez, E.~Fishbach, G.~L.~Klimchitskaya, D.~E.~Krause, and
V.~M.~Mostepanenko, Ann. Phys. \textbf{318}, 374 (2005); F.~Capasso, J.~Munday, D.~Iannuzzi and
H.B.~Chan, IEEE J. Sel. Top. Quant. Electr. \textbf{13}, 400 (2007).



\bibitem{Lamoreaux97} S.K.~Lamoreaux, Phys. Rev. Lett. {\bf 78} 5 (1997).
%repulsive hemispheres
\bibitem{Elizalde91} E.~Elizalde and A. Romeo, Am. J. Phys. {\bf 59}, 711 (1991).

\bibitem{CasimirPolder48} H. B. G. Casimir, and D. Polder,
%The Influence of Retardation on the London-van der Waals Forces,
Phys. Rev. {\bf 73}, 360 (1948).

\bibitem{Schmidt06} A. Gusso, A. G. M. Schmidt,
%Repulsive Casimir forces produced in rectangular cavities: Possible measurements and applications
Braz.J.Phys. {\bf 36}, 168 (2006).
%Cite as: arXiv:cond-mat/0410218v2
%cond-mat.other

\bibitem{Hushwater97} V. Hushwater, thesis, University of Maryland 1997,
%Radiation Pressure Approach to the Repulsive Casimir Force
arXiv:quant-ph/9909084.

\bibitem{Barton06} G.~Barton, Phys. Rev. {\bf D 73}, 065018 (2006).

\bibitem{Fulling07a} S.A.Fulling, L.~Kaplan and J.H.~Wilson, Phys. Rev. A{\bf 76},012118 (2007)
    [arXiv:quant-ph/0703248]; see also arXiv:quant-ph/0608122; Xiang-hua Zhai and Xin-zhou Li, Phys. Rev. D{\bf 76}, 047704 (2007)
    %\emph{Casimir Pistons with hybrid boundary conditions}
    [arXiv:hep-th/0612155v2].

\bibitem{Kenneth02} O. Kenneth, I. Klich, A. Mann and M.~Revzen, Phys. Rev. Lett. {\bf89}, 033001
    (2002).
% misleading 2-body approximation




















\bibitem{Schaden06b} M. Schaden, preprint \emph{Semiclassical Electromagnetic Casimir
    Self-Energies},
    arXiv:hep-th/0604119.










\bibitem{Boyer68} T.~H.~Boyer,
%``Quantum Electromagnetic Zero Point Energy Of A Conducting Spherical Shell
%And The Casimir Model For A Charged Particle,''
Phys.\ Rev.\  {\bf 174}, 1764 (1968).
%%CITATION = PHRVA,174,1764;%%





\bibitem{SW71} Stewartson K, Waechter R T.
% On hearing the shape of a drum: further results.
 {\it Proc Camb Philos Soc}{\bf 69} (1971)  353–363.

\bibitem{Svaiter92} N.F. Svaiter and B.F.Svaiter, J. Phys. A 25,979 1992.

\bibitem{Stroock93} D.W. Stroock,
Probability Theory, an Analytic View (Cambridge Univ. Press.,
    Cambridge 1993).

\bibitem{Goldman96}A. Goldman,
%includes spectral function relation to hull???
Probab. Theory Related Fields {\bf 105} (1996) 57.


\bibitem{littlejohn92} Littlejohn R.G.,
% The Van Vleck formula, Maslov theory, and phase space geometry,
 {\it J. Stat. Phys.} {\bf 68} (1992), 7--50.

\bibitem{kac69} Kac M.,
% Some mathematical models in science
 {\it Science} {\bf 166} (1969),  pp. 695–699.

\bibitem{Poincare05}Poincare, H.,
% Sur les lignes geodesiques des surfaces convexes,
 {\it Trans. Amer. Math. Soc.} {\bf 5} (1905), 237--274


% for waechter, interesting to see http://www.jstor.org/stable/52446?seq=29
% where a reference is made to Stewartson-Waechter Method (page 555)



\bibitem {Waechter72} Waechter R T.
% On hearing the shape of a drum: An extension to higher dimensions.
 {\it Proc. Camb. Philos. Soc.}, (1972) {\bf 72}, 439–447

\bibitem {Gelfand53} Gel'fand I M, Levitan P M.,
% On a simple identity for the eigenvalues of a second order differential operator.,
 {\it Dok Akad Nauk SSSR}, (1953), {\bf 88} 593–596

\bibitem {Gottlieb88} Gottlieb H P.,
% Eigenvalues of the Laplacian for rectilinear regions.,
 {\it J. Austral Math Soc Ser B}, (1988), {\bf 29}, 270–281

\bibitem{BD} Roger Balian, Bertrand Duplantier,
% Geometry of the Casimir Effect,
 {\it arXiv:quant-ph/0408124v1}

\bibitem{Power64} Power E.A.,
 Introductory Quantum Electrodynamics (London: Longman) 1964



\bibitem {Casimir48} Casimir H.B.G.,
% On the attraction between two perfectly conducting plates
 {\it Proc. Kon. Nederland. Akad. Wetensch.} {\bf B51}, 793 (1948)



% this is for stressing the connection between Brownian bridges
% and the partition function of the Laplacian
\bibitem {Lowther} Lowther, G.
% A Generalized Feynman-Kac Formula For One Dimensional Processes
 {\it arXiv:0803.3303v1 [math.PR]} (2008)

%% it is Oksendal, NOT Oksendahl!

\bibitem {Oksendal} Bernt Oksendal,
% Stochastic Differential Equations: An Introduction with Applications (Universitext),
 6th Edition, Springer, 2003

\bibitem{boyer0} Boyer T.H.,
% Van der Waals forces and zero-point energy for dielectric and
% permeable materials,
 {\it Phys. Rev.~A} {\bf 9} (1974), 2078--2084.

\bibitem {Boyer1} Boyer, Timothy H.,
% The Classical Vacuum
 {\it Scientific American} (1985)

\bibitem {Boyer2} Boyer, Timothy H.
% Random electrodynamics: The theory of classical electrodynamics with classical
%electromagnetic zero-point radiation,
 {\it Phys. Rev.} 790-808.

\bibitem{Brack} Brack M., Bhaduri R.K.,
 Semiclassical physics,
 Addison-Wesley, Reading, 1997

%% Mount Lectures?

\bibitem {Mateescu} M. Schaden and L. Mateescu,
% Weak and Repulsive Casimir Force in Piston Geometries,
 {\it [arXiv:0705.3435v1]}
}


\end{thebibliography}
\end{document}